\documentclass[12pt,a4]{iopart}
\usepackage{iopams}  

\usepackage{epsfig,rotating}
\usepackage{graphicx}
\usepackage{epstopdf,color}

\def\cm{cm$^{-1}$\,}
\newcommand{\bra}[1]{\left<#1\right|}
\newcommand{\ket}[1]{\left|#1\right>}
\newcommand{\nvl}[2]{\left|#1\right>\left<#2\right|}

\newcommand{\Eq}[1]{Eq.~(\ref{#1})}

\begin{document}

\title[Laser Control of Dissipative Two-Exciton Dynamics]{Laser Control of Dissipative Two-Exciton Dynamics in Molecular Aggregates}

\author{Yun-an Yan, Oliver K\"uhn}

\address{Institute of Physics, University of Rostock, D-18051 Rostock, Germany}
\ead{oliver.kuehn@uni-rostock.de}
\begin{abstract}
There are two types of two-photon transitions in molecular aggregates, that is, non-local excitations of two monomers and local double excitations to some higher excited intra-monomer electronic state. As a consequence of the inter-monomer Coulomb interaction these different excitation states are coupled to  each other. Higher excited intra-monomer states are rather short-lived due to efficient internal conversion of electronic into vibrational energy. Combining both processes leads to the annihilation of an electronic excitation state, which is a major loss channel for establishing high excitation densities in molecular aggregates.

Applying theoretical pulse optimization techniques to a Frenkel exciton model it is shown that the dynamics of two-exciton states in linear aggregates (dimer to tetramer) can be influenced by ultrafast shaped laser pulses. In particular, it is studied to what extent the decay of the two-exciton population by inter-band transitions can be transiently suppressed. Intra-band dynamics is described by a dissipative hierarchy equation approach, which takes into account strong exciton-vibrational coupling in the non-Markovian regime.
\end{abstract}

% Uncomment for Submitted to journal title message
%\submitto{\JPA}
% Comment out if separate title page not required
\maketitle

\section{Introduction}%-------------------------------------------------------------
Molecular aggregates continue to provide inspiration and challenges to experiment and theory \cite{thorwart10,thorwart11,may11}. In terms of a microscopic understanding of the dynamics of elementary excitations recent advances due to  time-resolved nonlinear spectroscopy have been tremendous \cite{mukamel09:553,ginsberg09:1352}. However, despite the success of laser pulse control in rather different areas of research \cite{SfbBook}, relatively little emphasis  has been put on exciton dynamics. On the experimental side, feedback laser control has been applied to manipulate the branching ratio between internal conversion (IC) and energy flow in the light-harvesting antenna of purple bacteria \cite{wohlleben05:850}. There has been a number of simulations for light-harvesting systems by May and Br\"uggemann et al. where the focus was put on the transient generation of a single localized excitation within the aggregate \cite{bruggemann04:10529,bruggemann04:573,brueggemann06_322,brueggemann07_372}; non-biological aggregates have received no attention so far.

The quantum dynamics of excitons is well founded within the Frenkel exciton approach, which starts from a classification of aggregate's excitation states in terms of the number of simultaneously excited monomers \cite{renger01:137,knoester03:1}. Under weak irradiation conditions only a single excitation will be present, but in nonlinear optical experiments or in the presence of strong irradiation multiple excited states play an important role. The organic building blocks of molecular aggregates have a multitude of electronically excited states, i.e. there is always a state S$_n$ such that the S$_0$-S$_1$ excitation energy approximately matches that of  a S$_1$-S$_n$ transition. As a consequence  one needs to distinguish between local double excitations (LDE) and nonlocal double excitations (NDE) where the two excitations are localized at different monomers. Note that the NDE should not be confused with bi-exciton states, i.e. bound states formed by two excitons in the presence of different permanent dipoles in the involved electronic states \cite{knoester03:1}. LDE and NDE can couple via the Coulomb interaction. The manifestation of this coupling in nonlinear spectroscopy has been investigated in Refs. \cite{knoester95:2780,kuhn96:8586,kuhn97:809}. The presence of  LDEs has important consequences for the exciton dynamics since it leads to exciton-exciton annihilation (EEA) by virtue of an intramolecular IC triggered by non-adiabatic electronic transitions. The presence of this process in molecular aggregates is well established \cite{sundstrom88:2754,trinkunas01:4167,bruggemann09_140} and various theoretical approaches exist \cite{malyshev99_117,renger00:807,brueggemann01_11391,bruggemann03:746,may09:10086}, although it must be stated that a first principles determination of the respective IC rates is still out of reach. 

Understanding exciton dynamics in aggregates is impossible without taking into account the effect of exciton-vibrational coupling \cite{may11,renger01:137,kuehn97_213}. Here one can distinguish between approaches which either account for all vibrations on the same footing, i.e. in terms of a heat bath, or treat a few active vibrational degrees of freedom explicitly, but coupled to the heat bath of the remaining modes. Needless to say, the latter approach is more demanding as the dimension of the density matrix increases rapidly with the number of explicit modes. Early investigations along these lines therefore have been restricted to molecular dimer models with one active vibrational coordinate per monomer \cite{kuhn96:99,renger96:15654,renger97:3406,renger97}. More recent approaches include a Green's function description of exciton dynamics \cite{roden09:044909}, a multi-configuration time-dependent Hartree simulation \cite{seibt09:13475} or the Frenkel excitonic polaron treatment \cite{spano10:429}. These methods cannot treat finite temperature effects due to the coupling to a further heat bath in a microscopic manner.

The coupling of the electronic degrees of freedom to a single or site-specific heat bath is usually treated using dissipation theory \cite{may11}. Recently, the so-called hierarchy equation (HE) method \cite{ishizaki05:3131,tanimura06_062001,ishizaki09_234111,zhu11_5678,shao04_5053,yan04_216,gelszinis11_245430} has enjoyed great popularity as it promises a non-perturbative and non-Markovian description of exciton dynamics, which is numerically equivalent, e.g. to the influence functional approach within a path integral formulation \cite{nalbach11_063040}. In the present context one should note that a distinction between different types of modes, i.e. low-frequency solvent modes versus high-frequency intramolecular vibrations, can be introduced via the spectral density. This is of particular relevance since the chromophores used to assemble artificial molecular aggregates usually show vibronic side bands in their absorption spectra pointing to the prominent role of intramolecular high-frequency vibrations \cite{ambrosek11_17649}.

The focus of the present contribution is on the laser control of dissipative two-exciton dynamics in linear aggregates. Thereby we take into account the effect of EEA and of a coupling to a heat bath being composed of an effective high-frequency mode as well as of a continuous distribution of low-frequency modes. For the solution of the density matrix equations of motion we will use a HE approach. Two-exciton populations are usually quenched by EEA. Since the latter is a local process we will ask the question whether an excitonic wave packet can be prepared such as to transiently suppress EEA by its composition in terms of LDE states. In the following Section \ref{sec:theory} we will outline the density matrix approach and its interface to optimal control theory (OCT). Section \ref{sec:results} will start with a discussion of the field free dynamics focussing on the comparison between the Markovian and non-Markovian limits. Subsequently, laser driven dynamics is discussed. The results are summarized in Sec. \ref{sec:sum}.

\section{Theory} %------------------------------------------------------------------
\label{sec:theory}
\subsection{Frenkel Exciton Hamiltonian}
Consider an aggregate composed  of $N$ monomers, each carrying three adiabatic electronic states ($a=g,e,f$) corresponding to the S$_0$, S$_1$, and some S$_n$ state. Thus we have the adiabatic electronic basis of local states $\ket{m,a}$,
$m=1,\ldots, N$. The electronic states of the aggregate can be classified as the zero excitation (ground) state \cite{may11}
\begin{eqnarray}
\ket{0} = \Pi_m \ket{m, g},
\end{eqnarray}
the singly-excited state
\begin{eqnarray}
\ket{m} = \ket{m, e}\Pi_{n\neq m} \ket{n, g},
\end{eqnarray}
the doubly-excited (LDE and NDE) states
\begin{eqnarray}
\label{eq:de2}
\ket{mm} &=& \ket{m, f}\Pi_{n\neq m} \ket{n, g}, \\
\label{eq:de11}
\ket{mn} &=& \ket{m, e}\ket{n,e}\Pi_{k\neq m,n} \ket{k, g}.
\end{eqnarray}
Restricting ourselves to these excitation the completeness relation reads
\begin{eqnarray}
1 = \ket{0}\bra{0} + \sum_m(\nvl{m}{m}+\nvl{mm}{mm})
    + \frac{1}{2}\sum_{m\neq n} \nvl{mn}{mn}.
\end{eqnarray}
The electronic Hamiltonian can be written as  
\begin{eqnarray}
\label{eq:hel}
\hat{H}_{\rm el} =  \hat{H}_{\rm ex} + \hat{H}_{\rm field}(t) + \hat{H}_{\rm IC} \, ,
\end{eqnarray}
 with the bare exciton Hamiltonian, 
 \begin{eqnarray}
\hat{H}_{\rm ex} =\hat{H}^{(0)}_{\rm ex}+ \hat{H}^{(1)}_{\rm ex} + \hat{H}^{(2)}_{\rm ex} \, ,
\end{eqnarray}
being composed of the ground state part
\begin{eqnarray}
 \hat{H}^{(0)}_{\rm ex} =  E_0 \ket{0}\bra{0} \, ,
\end{eqnarray}
where $E_0$ is the electronic ground state energy,
the single-exciton Hamiltonian
\begin{eqnarray}
\hat{H}^{(1)}_{\rm ex} &=& \sum_{m,e} E_{m,e} \nvl{m}{m} 
                  + \sum_{m\neq n} J_{mn} \nvl{m}{n} \, ,
\end{eqnarray}
and the two-exciton Hamiltonian
\begin{eqnarray}
\fl
\hat{H}^{(2)}_{\rm ex} = \sum_{m,f} E_{m,f}\nvl{mm}{mm}
	 + \frac{1}{2}\sum_{m\neq n} (E_{m,e}+E_{n,e}) \nvl{mn}{mn} \nonumber \\
 + \sum_{m\neq n}\sum_{k\neq m,n} J_{nk} \nvl{mn}{mk} 
	 + \sum_{m\neq n} \left(J_{mn}^{\rm (ef)} \nvl{mn}{mm}  + {\rm h.c.}\right) \, .
\end{eqnarray}
Here, $E_{m,e}$ and $E_{m,f}$ are the energies of electronic excitation at site $m$ for S$_0$-S$_1$ and S$_0$-S$_n$, respectively. Further, $J_{mn}$ is the coupling between site $m$ and site $n$ leading to single exciton transfer and 
$J^{(\rm ef)}_{mn}$ is the Coulomb coupling responsible for the two-exciton dynamics.

It is customary to introduce Frenkel exciton eigenstates which follow from the diagonalization of the exciton Hamiltonian
\begin{eqnarray}
\hat{H}_{\rm ex}\ket{\alpha} = E_\alpha\ket{\alpha}.
\end{eqnarray}
The eigenstates can be decomposed in terms of local excitation states,
$\ket{a} (a=0,\,m,\,mm,\,mn)$,  as follows
\begin{eqnarray}
\ket{\alpha} = \sum_a C_{a}(\alpha)\ket{a}.
\end{eqnarray}
The eigenstates separate into $\mathcal{M}$-exciton manifolds (here $\mathcal{M}=0,1,2$).
Frequently, we will also use a notation where the states in the $\mathcal{M}$-exciton manifolds are counted according to increasing energy, i.e. $\ket{\mathcal{M}_k}$.

In Eq. (\ref{eq:hel}) the aggregate is assumed to interact with an external laser field treated in dipole approximation, i.e.
\begin{eqnarray}
\hat{H}_{\rm field}(t) =  - E(t) \hat{\mu} \, ,
\end{eqnarray}
where the field $E(t)$ is oriented parallel to the transition dipoles. In the local Frenkel exciton basis 
the dipole operator for the aggregate reads
\begin{eqnarray}
\hat{\mu} = \sum_m \left( \mu_{m,e}\nvl{m}{0}
       + \mu_{m,f} \nvl{mm}{m} 
       + \sum_{n\neq m}\mu_{n,e}\nvl{mn}{m}\right) + {\rm h.c.}.
\end{eqnarray}
Here, $\mu_{m,e}(\mu_{m,f})$ is the transition dipole for S$_0$-S$_1$(S$_1$-S$_n$) transition of monomer $m$.

Finally, we consider the IC process described by $\hat{H}_{\rm IC}$ in Eq. (\ref{eq:hel}). IC has its origin in the breakdown of the Born-Oppenheimer approximation, e.g. at conical intersections. The respective non-adiabaticity operator triggering transitions between adiabatic electronic states is proportional to the momentum operator $\hat{P}_{m,\xi}$ of the involved nuclear degrees of freedom (coordinates $\hat{Q}_{m,\xi}$) counted by the index $\xi$ at monomer $m$.  The non-radiative life time of the $S_1$ state is usually in the nanosecond range. Therefore we will restrict ourselves to the IC between adiabatic states $\ket{m,f}$  and $\ket{m,e}$. Hence the coupling becomes
\begin{eqnarray}
\label{eq:hic}
\hat{H}_{\rm IC} &=& \sum_m \hat{\Pi}_m^{\rm (IC)}\sum_\xi  g^{\rm (IC)}_{m,\xi} \hat{P}_{m,\xi} \, , \nonumber \\ 
\hat{\Pi}_m^{\rm (IC)}     &=& \nvl{m}{mm} + \nvl{mm}{m} \, .
\end{eqnarray}
In the following we will assume that  the  coupling strength  is site-independent, i.e. $g^{\rm (IC)}_{m,\xi}=g^{\rm (IC)}_{\xi}$
\subsection{Exciton-Vibrational Coupling: System-Bath Model} %--------------------------------------
\label{sec:evc}
Exciton-vibrational coupling leading to intra-band phase and energy relaxation is introduced in the spirit of the 
system-bath approach, i.e. we have 
\begin{eqnarray}
\label{eq:hsb}
\hat{H} & = & \hat{H}_{\rm S} + \hat{H}_{\rm B} + \hat{H}_{\rm S-B}  \nonumber\\
 & = & \hat{H}_{\rm el} + \hat{H}_{\rm vib} + \hat{H}_{\rm el-vib} \, .
\end{eqnarray}
The bath modes are treated in  harmonic approximation, i.e. $\hat{H}_{\rm vib}$ is the standard harmonic oscillator Hamiltonian. Concerning the system-bath coupling we will distinguish between two types of modes (see also Ref. \cite{polyutov12_21}). First, local high-frequency modes, $\{q_{m,\xi}\}$, which usually give rise to a vibrational progression in the monomer absorption spectrum \cite{ambrosek11_17649}. Second, global low-frequency modes, $\{x_{\xi}\}$, contributing via a continuous spectrum. Hence we have
\begin{eqnarray}
\hat{H}_{\rm el-vib} & = & \sum_m \hat{H}_m^{\rm (H)} + \hat{H}^{\rm (L)}\,.
\end{eqnarray}
For the coupling to the local high-frequency modes we will use the model Hamiltonian
\begin{eqnarray}
\label{eq:evc1}
\hat{H}_{m}^{\rm (H)}&= &\sum_\xi \hat{q}_{m,\xi} \Big[
   g^{\rm (H)}_{m,\xi,e}\nvl{m}{m} + g^{\rm (H)}_{m,\xi,f} \nvl{mm}{mm} \nonumber\\
&+& 
    \sum_{n\neq m} g^{\rm  (H)}_{m,\xi,e} \nvl{mn}{mn}\Big] \, .
\end{eqnarray}
In the following we will assume that the coupling is the same for all monomers, i.e. $g^{\rm (H)}_{m,\xi,a}=g^{\rm (H)}_{\xi,a}$. Further, we will use $ g^{\rm (H)}_{\xi}= g^{\rm (H)}_{\xi,e}$
and $g^{\rm (H)}_{\xi,f} = \kappa g^{\rm (H)}_{\xi}$
\cite{kuhn97:4154}. With these assumptions \Eq{eq:evc1} can be written as
\begin{eqnarray}
\label{eq:hsb}
\hat{H}_{m}^{\rm (H)}&=&\hat{h}_m
              \sum_\xi g^{\rm (H)}_{\xi} \hat{q}_{m,\xi},\nonumber\\
\hat{h}_m &=&\nvl{m}{m} + \kappa \nvl{mm}{mm} +  \sum_{n\neq m}\nvl{mn}{mn} \, .
\end{eqnarray}
For the coupling to the low-frequency modes we will invoke the same approximations and arrive at the Hamiltonian
\begin{eqnarray}
\hat{H}^{\rm (L)}&=&\hat{h}^{\rm (tot)} \sum_\xi g^{\rm (L)}_{\xi} \hat{x}_\xi,\nonumber \\
\hat{h}^{\rm (tot)}&=& \sum_m \hat{h}_m.
\end{eqnarray}

\subsection{Dissipation Models}
\subsubsection{Internal Conversion}
The IC rate between S$_n$ and S$_1$ states is rather large for typical chromophores, reaching time scales on
the order of about 100 fs \cite{kuhn11_47}. On this time scale the S$_n$ population decays into the S$_1$ state, thereby passing many electronic states (for the case of perylene bisimides $n$ would be of the order of about 30 \cite{ambrosek11_17649}). Hence it is justified to assume that the memory time associated with the IC process is even shorter than 100 fs and EEA can be treated in Markovian approximation. Treating $\hat{H}_{\rm IC}$ in \Eq{eq:hic}  as a perturbation of  the bare exciton Hamiltonian within second order perturbation theory one can write the contribution to the Quantum Master Equation for the reduced exciton density operator $\hat{\rho}$  in terms of 
the Lie operator $\mathcal{R}^{\rm (IC)}$ which operates on $\hat{\rho}$ as \cite{may11}
\begin{eqnarray}
\label{eq:redic}
i\hbar\mathcal{R}^{\rm (IC)}\hat{\rho}= - \sum_m \Big[\hat{\Pi}_m^{\rm (IC)}, (\hat{\Xi}_m \hat{\rho}
                                  - \hat{\rho} \hat{\Xi}_m^\dagger)\Big] \, .
\end{eqnarray}
Here $\hat{\Xi}_m$ is defined as
\begin{eqnarray}
\label{eq:redxi}
\hat{\Xi}_m = \int^\infty_0 dt \, \alpha_{m}^{\rm (IC)}(t)
             \exp(-i\hat{H}_{\rm ex}t/\hbar)\hat{\Pi}_m^{\rm (IC)}\exp(i\hat{H}_{\rm ex}t/\hbar) \, 
\end{eqnarray}
and  $\alpha_{m}^{\rm (IC)}(t)$ is correlation function of the part $\sum_\xi  g^{\rm (IC)}_{m,\xi} \hat{P}_{m,\xi}$ of \Eq{eq:hic} (note that we assume that the momenta at different sites are uncorrelated).
By using the completeness relation for the eigenstates, one has
\begin{eqnarray}
\hat{\Xi}_m = \sum_{\alpha,\beta}\hat{\Pi}^{\rm (IC)}_{m,\alpha\beta}
    J_m^{\rm (IC)}\left((E_\beta-E_\alpha)/\hbar\right) \nvl{\alpha}{\beta} \, .
\end{eqnarray}
Here, $J_m^{\rm (IC)}(\omega) = \int^\infty_0dt  \exp(i\omega t) \alpha_{m}^{\rm (IC)}(t)$
is the Fourier transformation of the correlation function and $\Pi^{\rm (IC)}_{m,\alpha\beta} = 
\bra{\alpha}\hat{\Pi}_m^{\rm (IC)}\ket{\beta}$. For simplicity the Debye spectra is used below
\begin{equation}
\label{eq:Jic}
J_{m}^{\rm (IC)}(\omega) = \eta_{\rm IC} \omega \frac{\Lambda_{\rm IC}^2}{\omega^2 + \Lambda_{\rm IC}^2}
\end{equation}
\subsubsection{Separation of Time Scales in the Response Function of the Oscillator Bath}
\label{sec:timescale}
In general the influence of the bath degrees of freedom due to exciton-vibrational coupling cannot be treated in Markovian approximation.  It is fully characterized by  the response function, $\alpha(t)$, of the bath, which in turn is 
determined by the spectral density function $J(\omega)$ 
\begin{eqnarray}
\label{eq:resp}
\alpha(t) &=& \frac{1}{\pi} \int^\infty_0d\omega J(\omega) 
   \left[\coth(\beta\hbar\omega/2) \cos(\omega t) -i \sin(\omega t )\right],
  \nonumber \\
J(\omega) &=& \frac{\pi}{2}\sum_\xi \frac{g_\xi^2}{\mu_\xi\omega_\xi}
      \delta(\omega - \omega_\xi).
\end{eqnarray}
Here, $g_\xi, \mu_\xi$, and $\omega_\xi$ are the coupling constant, reduced mass, and frequency, respectively, of the $\xi$th oscillator and $\beta=1/k_{\rm B}T$.
In the following we will specify the spectral density to the coupling models discussed in Section \ref{sec:evc}.

First, we consider the case of the local high-frequency mode, \Eq{eq:evc1} for which we assume a damped Brownian oscillator model to hold. This can be described by the following spectral density \cite{grabert88_115}
\begin{eqnarray}
\label{eq:jhigh}
J_{\rm H}(\omega) = 
  \frac{\eta_{\rm H}\omega_0^3\omega\Lambda_{\rm H}}{(\omega^2 - \omega_0^2)^2 + \Lambda_{\rm H}^2\omega^2} \, .
\end{eqnarray}
Below we will assume the underdamped limit, where the central mode frequency $\omega_0$ is much larger than 
the cutoff $\Lambda_{\rm H}$. For this spectral density the response function can be expanded  into a sum of exponentials as follows
\begin{eqnarray}
\label{eq:high}
\fl
\alpha_{\rm H}(t) = \frac{\eta_{\rm H}\hbar\omega_0^3}{4\xi}
    \left\{-\exp(\Omega_1 t) \left[\coth(i\hbar\beta\Omega_1/2) - 1\right]
           +\exp(\Omega_2 t) \left[\coth(i\hbar\beta\Omega_2/2) - 1\right]
    \right\} \nonumber \\
 - 2 \eta_{\rm H}\omega_0^3\Lambda_{\rm H}^2\hbar \beta \sum_{k=1}^\infty
     \frac{\nu_ke^{-\nu_k t}}{(\nu_k^2 + \omega_0^2)^2 - \Lambda_{\rm H}^2\nu_k^2} 
           \nonumber \\
  \equiv \sum_{k=1}^2 b_k e^{-\Omega_k t}
   + \sum_{k=1}^{N_{\rm H}} c^{\rm (H)}_k \exp(-\nu_k t) + \alpha_{\rm H}^{\rm (short)}(t).
\end{eqnarray}
where $\Omega_{1,2} = \Lambda_{\rm H}/2 \pm i \xi$, $\xi = \sqrt{\omega^2_0 - \Lambda_{\rm H}^2/4}$.

In \Eq{eq:high}, $\nu_k = 2\pi k/(\hbar \beta) (k>0)$ is the $k$th Matsubara frequency of the 
bath, and $N_{\rm H}$ is the smallest integer satisfying 
$\nu_{N_{\rm H}+1} \gg \Lambda_{\rm H}$. As discussed in Ref.~\cite{ishizaki05:3131} the response function can be split into two parts: a long memory contribution (the first two terms in the last line of  
Eq.~(\ref{eq:high})) and a short memory part $\alpha_{\rm H}^{\rm (short)}(t)$. 
The latter one contains all the terms with short memory in the 
Matsubara summation and will be treated in Markovian approximation  within the hierarchy equations to be discussed below.

The low-frequency bath will be described by the Debye spectral density
\begin{eqnarray}
\label{eq:jlow}
J_{\rm L}(\omega) = \eta_{\rm L} \omega \frac{\Lambda_{\rm L}^2}{\omega^2 + \Lambda_{\rm L}^2}
\end{eqnarray}
where $\eta_{\rm L}$ is the coupling strength and $\Lambda_{\rm L}$ is the 
frequency cutoff of the bath. The response function in Eq. (\ref{eq:resp}) then 
can be expressed as follows
\begin{eqnarray}
\label{eq:low}
\fl
\alpha_{\rm L}(t) =
\frac{\eta_{\rm L}\Lambda_{\rm L}^2}{2}
     \left[\cot\left(\beta\hbar\Lambda_{\rm L}/2\right) - i\right] e^{-\Lambda_{\rm L} t} 
%%%   \nonumber \\ && 
    +  \frac{2 \eta_{\rm L}\Lambda_{\rm L}^2}{\hbar \beta}
         \sum_{k=1}^\infty\frac{\nu_ke^{-\nu_k t}}{\nu_k^2 - \Lambda_{\rm L}^2} \nonumber \\
  \equiv 
  \sum_{k=0}^{N_{\rm L}} c^{\rm (L)}_k \exp(-\nu_k t) + \alpha_{\rm L}^{\rm (short)}(t),
\end{eqnarray}
where $\nu_0 = \Lambda_{\rm L}$, $\nu_k = 2\pi k/(\hbar \beta) (k>0)$ is the $k$th Matsubara frequency of the 
bath, and $N_{\rm L}$ is the smallest integer satisfying 
$\nu_{N_{\rm L}+1} \gg \Lambda_{\rm L}$. The same time scale separation has been introduced as for the high-frequency bath.%
\subsection{Hierarchy Equations for the Dissipative Multi-exciton Dynamics}
In order to describe the non-Markovian and non-perturbative exciton dynamics we will utilize a HE approach to propagate the reduced exciton density matrix in eigenstate representation. Specifically we have employed the stochastic decoupling procedure due to Shao and coworkers which is briefly sketched in the Appendix \cite{shao04_5053,yan04_216}. Since the HE approach starts from the influence functional~\cite{ishizaki05:3131,xu05_041103},
it is based on the fact that the response function for bath is written as a sum of exponentials (see previous section).

The resulting HEs of motion for the dissipative exciton dynamics of the aggregate are given by
\begin{eqnarray}
\label{eq:EOM}
\fl
i\hbar \frac{\partial}{\partial t}\hat{\rho}_{ABV}(t) 
    = - i\hbar \left(\sum_{m=1}^{N}\sum_{j=1}^{2} A_{mj} \Omega_{j} 
          + \sum_{m=1}^{N}\sum_{j=1}^{N_{\rm H}} B_{mj}\nu_j
          + \sum_{k=0}^{N_{\rm L}} V_j \nu_j + \mathcal{R}^\prime \right) 
               \hat{\rho}_{ABV}(t) \nonumber \\
    + \left[\hat{H}_{\rm ex}+\hat{H}_{\rm field}(t), \hat{\rho}_{ABV}(t)\right] 
    + \sum_{j=0}^{N_{\rm L}}\nu_j^{-1}\sqrt{(V_j+1)|c_j^{\rm (L)}|} \left[\hat{h}^{\rm (tot)}, \hat{\rho}_{ABV^+_j}(t)\right] \nonumber \\
    + \sum_{m=1}^{N}\sum_{j=1}^{2}\Lambda_{\rm H}^{-1}\sqrt{(A_{mj}+1)|c_0^{\rm (H)}|}\left[\hat{h}_m, \hat{\rho}_{A^+_{mj}BV}(t)\right] \nonumber \\
    + \sum_{m=1}^{N}\sum_{j=1}^{N_{\rm H}}\nu_j^{-1}\sqrt{(B_{mj}+1)|c_j^{\rm (H)}|}\left[\hat{h}_m, \hat{\rho}_{AB^+_{mj}V}(t)\right] \nonumber \\
    + \hbar\sum_{m=1}^{N}\sum_{j=1}^{2}\Lambda_{\rm H}\sqrt{\frac{A_{mj}}{|c_0^{\rm (H)}|}}
          \left(\alpha_{j;+}\left[ \hat{h}_m, \hat{\rho}_{A^-_{mj}BV}(t)\right]
        - \alpha_{j;-}\left\{\hat{h}_m, \hat{\rho}_{A^-_{mj}BV}(t)\right\}\right) \nonumber \\
    + \hbar \sum_{m=1}^{N}\sum_{j=1}^{N_{\rm H}}\nu_j\sqrt{\frac{B_{mj}}{|c_j^{\rm (H)}|}}\left[\hat{h}_m, \hat{\rho}_{AB^-_{mj}V}(t)\right]
       + \hbar \sum_{j=1}^{N_{\rm L}}\nu_j\sqrt{\frac{V_{j}}{c^{\rm (L)}_{j}}} \left[ \hat{h}^{\rm (tot)}, \hat{\rho}_{ABV^-_{j}}(t)\right] \nonumber \\
     +  \Lambda_{\rm L}\sqrt{\frac{V_{0}}{|c^{\rm (L)}_0|}}\left(i\hbar c^{\rm (L)}_{0,i}\left\{\hat{h}^{\rm (tot)}, \hat{\rho}_{ABV^-_{j}}(t)\right\} 
              +   c^{\rm (L)}_{0,r}\left[\hat{h}^{\rm (tot)}, \hat{\rho}_{ABV^-_{j}}(t)\right]\right)
\end{eqnarray}
where $A$ is an $N\times 2$ integer matrix, 
  with its matrix element $A_{mj}$ corresponding the contribution 
  from $\Omega_j$ in $m$-th molecule (Eq.~(\ref{eq:high})). 
$B$ is an $N\times N_{\rm H}$ integer matrix with its matrix 
  element $B_{mj}$ counting the $j$-th Matsubara frequency 
  in the $m$-th molecule (Eq.~(\ref{eq:high})).
And $V$ is an $N_{\rm L}+1$ dimensional integer vector.
  $V_0$ refers to the cutoff frequency and $V_j$ ($j\neq 0$) to 
  the $j$-th Matsubara frequency in the low frequency bath (Eq.~(\ref{eq:low})).
The operator $``+/-"$ is defined as
$\left[M^\pm_{mj}\right]_{nl} = M_{nl} \pm \delta_{mn} \delta_{jl}$ 
for the matrix and
$\left[V^\pm_j\right]_k = V_{k} \pm \delta_{kj}$ for the vector.

$\mathcal{R}^\prime$ is the Redfield super-operator accounting for the 
exciton-exciton annihilation and the short memory effects of the 
high and low frequency environments, $[\cdot,\,\cdot]$ denotes 
the commutator and $\{\cdot,\,\cdot\}$ denotes the anti-commutator.
For the definition of $\alpha_{k,\pm}$ and $c^{\rm (H)}_0$, see Appendix.

In the above equations, the first term $\hat{\rho}_{\textbf{000}}$ is 
the reduced density matrix for the exciton and the other elements reflect
the finite memory effect due to the exciton-vibration coupling.
The initial condition for the hierarchy is 
$\hat{\rho}_{\textbf{000}}(0) = \hat{\rho}(0)$ 
and $\hat{\rho}_{ABV}(0) = 0$ if any of the elements of A, B and V
is nonzero.

In the Markovian limit, the HEs are reduced to the Quantum Master Equation
\begin{eqnarray}
\label{eq:qme}
 i\hbar \frac{\partial}{\partial t}\hat{\rho} &=& \left[\hat{H}_{\rm ex}, \hat{\rho}\right] 
           - E(t) \left[\hat{\mu},\hat{\rho}\right]
           - i\hbar\mathcal{R}\hat{\rho},
\end{eqnarray}
where $\mathcal{R}$ is the Redfield superoperator
\begin{eqnarray}
i\hbar \mathcal{R}\hat{\rho} &=& i\hbar \mathcal{R}^{\rm (IC)}\hat{\rho} 
   - \left[\hat{h}^{\rm (tot)},  (\hat{\Xi}^{\rm (tot)} \hat{\rho} 
       - \hat{\rho} \hat{\Xi}^{{\rm (tot)}\dagger}) \right] \nonumber \\
   &-& \sum_m^N\left[\hat{h}_m,  (\hat{\Xi}_m \hat{\rho} - \hat{\rho} \hat{\Xi}_m^\dagger) \right].
\end{eqnarray}
Here $\mathcal{R}^{\rm (IC)}$ is the superoperator accounting for the IC as defined in 
Eq.~(\ref{eq:redic}). $\hat{\Xi}^{\rm (tot)}$ and $\hat{\Xi}_m$ are the dissipative 
operators defined in the similar way as Eq.~(\ref{eq:redxi}) for the low frequency 
and high frequency bath in $m$-th molecule, respectively.
\subsection{Optimal Control Theory}
\label{sec:oct}
For the design of laser fields for  driving the exciton dynamics we will employ optimal control theory (see, eq. Ref. \cite{brif10:075008}; the present implementation follows Ref. \cite{may11}). Here the goal is to find a laser field $E(t) $ such as to optimize a target at a certain time $t=T$. This can be cast into an optimization problem for the functional
\begin{eqnarray}
\label{eq:octfunc}
{\mathcal J}(E; T) = \textrm{Tr}\{\hat{O} \hat{\rho}(T)\} 
- \frac{\lambda}{2}\left(\int_0^{T}\frac{|E(t)|^2}{\sin^2(\pi t/T)} dt - I_0\right),
\end{eqnarray}
where $\hat{O}$ is the projection operator onto the  target state,  $\lambda$ is 
the penalty factor for strong fields. In this work $\lambda$ is treated as a Lagrangian 
multiplier such as to keep the integrated intensity close to $I_0$.
To simplify matters we will not employ the HE approach at this point but 
resort to the Markovian approximation, which leads to a set of two coupled equations.

 Optimizing the functional ${\mathcal J}$ with  respect to the field one gets the expression
 \begin{eqnarray}
\label{eq:Eoct}
 E(t) = \frac{i}{\lambda\hbar}\textrm{Tr}\{\hat{\sigma}(t)[\hat{\mu}, \hat{\rho}(t)]\} \sin^2(\pi t/T),
 \end{eqnarray}
 where $\hat{\sigma}(t)$ is an auxiliary operator propagating backward in time,
 starting from $\hat{O}$ at $t=T$ with the following differential equation 
\begin{eqnarray}
\label{eq:sigma}
i\hbar \frac{\partial}{\partial t} \hat{\sigma} &=& \left[\hat{H}_{\rm ex}, \hat{\sigma}\right] 
          - E(t) [\hat{\mu},\hat{\sigma}]
%         - \frac{i}{\lambda\hbar}\textrm{Tr}\{\hat{\sigma}[\hat{\mu},\hat{\rho}]\}
            + i\hbar\tilde{\mathcal{R}}\hat{\sigma}, \nonumber \\
i\hbar \tilde{\mathcal{R}} \hat{\sigma} &=& 
\sum_m^N \left(\hat{\Xi}^{\rm (IC)\dagger}_m\left[\Pi^{\rm (IC)}_m, \hat{\sigma}\right]
- \left[\Pi^{\rm (IC)}_m, \hat{\sigma}\right]\hat{\Xi}^{\rm (IC)}_m\right) \nonumber \\
&+& \sum_m^N \left(\hat{\Xi}^{\dagger}_m\left[\hat{h}_m, \hat{\sigma}\right]
- \left[\hat{h}_m, \hat{\sigma}\right]\hat{\Xi}_m\right) \nonumber \\
&+&  \hat{\Xi}^{\rm (tot)\dagger}\left[\hat{h}^{\rm (tot)}, \hat{\sigma}\right]
- \left[\hat{h}^{\rm (tot)}, \hat{\sigma}\right]\hat{\Xi}^{\rm (tot)} \, .
\end{eqnarray}
Eqs.~(\ref{eq:qme}) and (\ref{eq:sigma}) are nonlinear equations coupled through 
Eq.~(\ref{eq:Eoct}), which can be solved iteratively. 

The fields obtained by the OCT  equations will be characterized by their  XFROG (cross-correlation frequency-resolved optical gating \cite{linden98_119})  trace defined as 
\begin{eqnarray}
\label{eq:xfrog}
I(\omega,t)= \left| \int dt' E( t^\prime) G( t^\prime-t)e^{-i\omega t^\prime} \right|^2,
\end{eqnarray}
where $G(t)$ is the gate function having the form
\begin{eqnarray}
G(t) = \frac{1}{2} ( \textrm{erf}[2(t+\tau/2)/\Delta] + \textrm{erf}[2(t-\tau/2)/\Delta]) \, .
\end{eqnarray}
In the above equation $\tau$ is the width of the gate and $\Delta$ is the width
of the shoulder.
\begin{figure}
\begin{center}
\includegraphics[width=\textwidth]{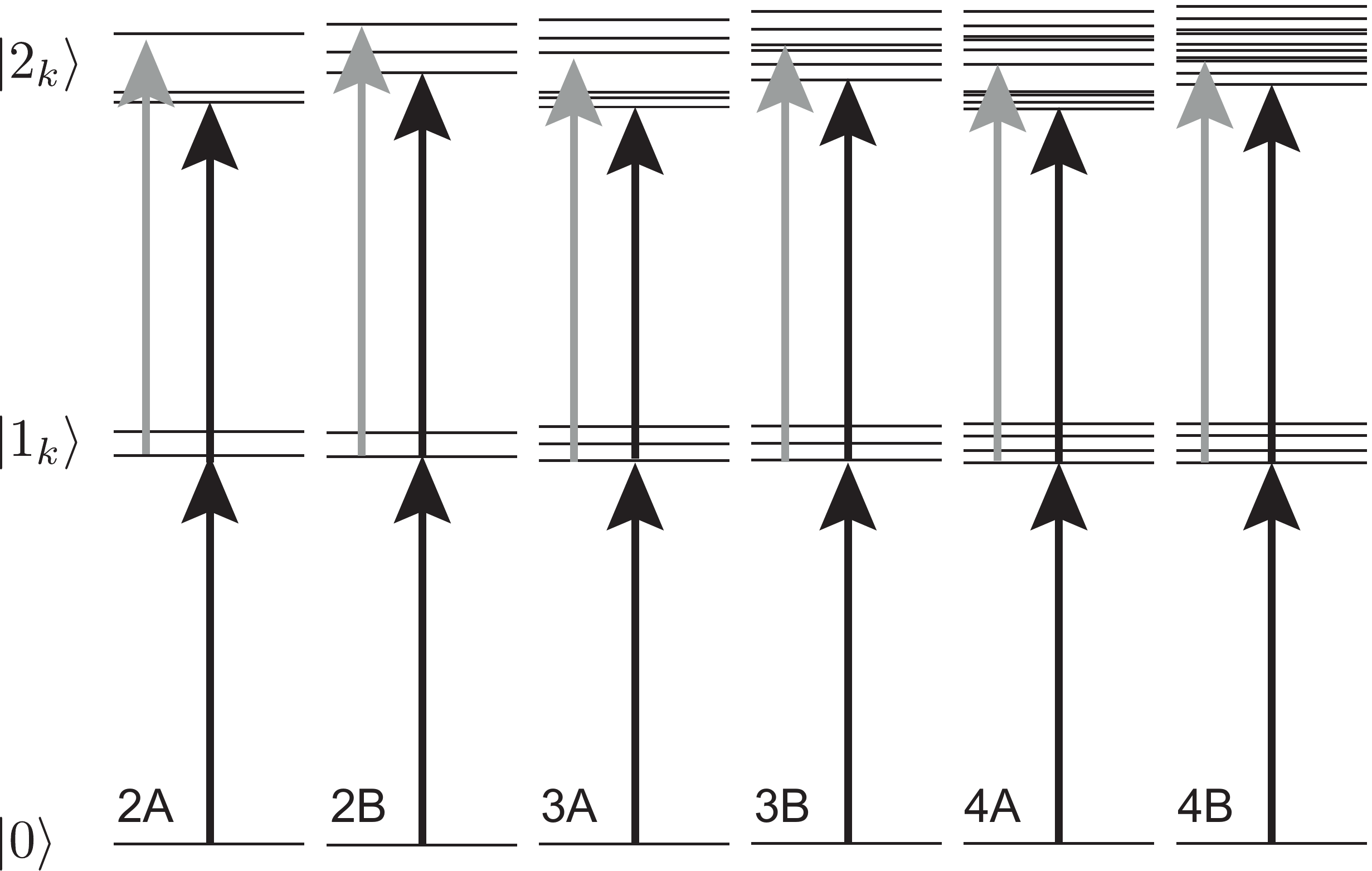}
\caption{Level scheme for eigenstates of the consider model systems $NA$ and $NB$. Black arrows mark the dominant transitions, grey arrows the next to dominant transitions starting from $\ket{1_1}$.}
\label{fig:level}
\end{center}
\end{figure}
\section{Results} %------------------------------------------------------------
\label{sec:results}
\subsection{Parameters}
We will apply the formalism outlined in Section \ref{sec:theory} to a model mimicking the situation in perylene bisimide aggregates. In particular we consider the case of $N$,$N$'-Di($N$-(2-aminoethyl)-benzamide)-1,6,7,12-tetra(4-tert-butylphenoxy)-3,4:9,10-perylenebiscarboximide (PBI) which is known to form J-aggregates \cite{li08:8074}. The photophysical properties of PBI monomers have been studied in Ref. \cite{ambrosek11_17649}. As far as the present model is concerned the following properties will be used: The S$_0$-S$_1$ transition energy is $E_{m,e}=2.13$ eV. The monomer S$_0$-S$_1$ transition dipole moment is 3.34 ea$_0$.

This transition is coupled to an effective vibrational mode of frequency 1415 \cm with a Huang-Rhys factor of 0.44.  Further we used $\kappa=1$ in Eq. (\ref{eq:hsb}). While the effective mode captured the vibrational side band observed in the experiment, it could not account for the general line broadening, which amounts to 1110 \cm thus indicating substantial coupling to a continuous distribution of low-frequency bath modes.  Hence, for the local high-frequency mode we take in the spectral density, Eq. (\ref{eq:jhigh})  $\omega_0 = 1415$~\cm and $\eta_{\rm H} = 0.44$, together with a phenomenological broadening width $\Lambda_{\rm H} = 100$ \cm. For the low-frequency heat bath consisting of intermolecular and solvent vibrations, a frequency cut-off of 
$\Lambda_{\rm L} = 100$ \cm is used in Eq. (\ref{eq:jlow}) together with a coupling strength of $\eta_{\rm L} = 2.0$, chosen such as  to yield considerable line broadening. The simulation is carried out at a temperature of  300 K, i.e. we can use $N_{\rm H}=N_{\rm L}=1$ in Eq. (\ref{eq:EOM}). Convergent results have been obtained for a hierarchy  order of seven.

The situation is more complicated for the monomeric excited state absorption (ESA). On the experimental side, the ESA spectrum has been obtained from a pump-probe spectrum, recorded after equilibration in the S$_1$ state. Since the procedure requires knowledge about ground state bleach and stimulated emission spectra at a detail which cannot be obtained, the extracted ESA spectrum may contain spurious features. On the theoretical side calculating highly excited states for a molecule as large as PBI is a challenge, in particular because standard time-dependent density functional theory will fail to describe transitions of double excitation character. In Ref. \cite{ambrosek11_17649} we applied the multireference variant of density functional theory to a reduced PBI system. Combining experimental and theoretical data it can be stated that (i) the excited state responsible for the S$_1$-S$_n$ transition is at $n> 20$. In other words, the internal conversion down to the S$_1$ state proceeds via a dense manifold of electronic states, whose microscopic description is out of reach. This suggests using a description in terms of an effective internal conversion rate as outlined in Section \ref{sec:theory}. For the actual value we have used an inverse rate of 100 fs. Within the spectral density model, Eq. (\ref{eq:Jic}), this amounts to choosing $\Lambda_{\rm IC}$=1000 \cm and $\eta_{\rm IC}=0.4$. (ii) Since reliable information on the transition dipole moment $\mu_{m,f}$ are not available we will use the value obtained within the harmonic oscillator picture, i.e. $\mu_{m,f}=\sqrt{2}\mu_{m,e}$ \cite{kuhn96:8586}. (iii) Concerning the anharmonicity $\Delta_m=E_{m,f}-2E_{n,e}$ we first note that in the range where 
$E_{m,f} \approx 2E_{n,e}$ two transitions have been observed. Since the magnitude of $\Delta_m$ decides about the mixing between local and nonlocal double excitation states \cite{kuhn96:8586} we will consider two cases. In case A the anharmonicity is large, $\Delta_m = -0.26$ eV, whereas in case B it is small, $\Delta_m = -0.04$ eV.

In Ref. \cite{ambrosek11_17649} a PBI monomer has been investigated only. The Coulomb coupling between adjacent monomers in the PBI aggregate has been estimated from the experimentally observed absorption line shifts upon aggregation. It has also been calculated in Ref. \cite{ambrosek_xx}. Below we will use the calculated value of 
$J_{mn}=-515$ \cm and only nearest neighbor coupling will be considered. The coupling between excited state transitions is not known and we will again use the harmonic oscillator approximation giving $J_{mn}^{\rm (ef)} = \sqrt{2} J_{mn}$.

\begin{figure}
\begin{center}
\includegraphics[width=\textwidth]{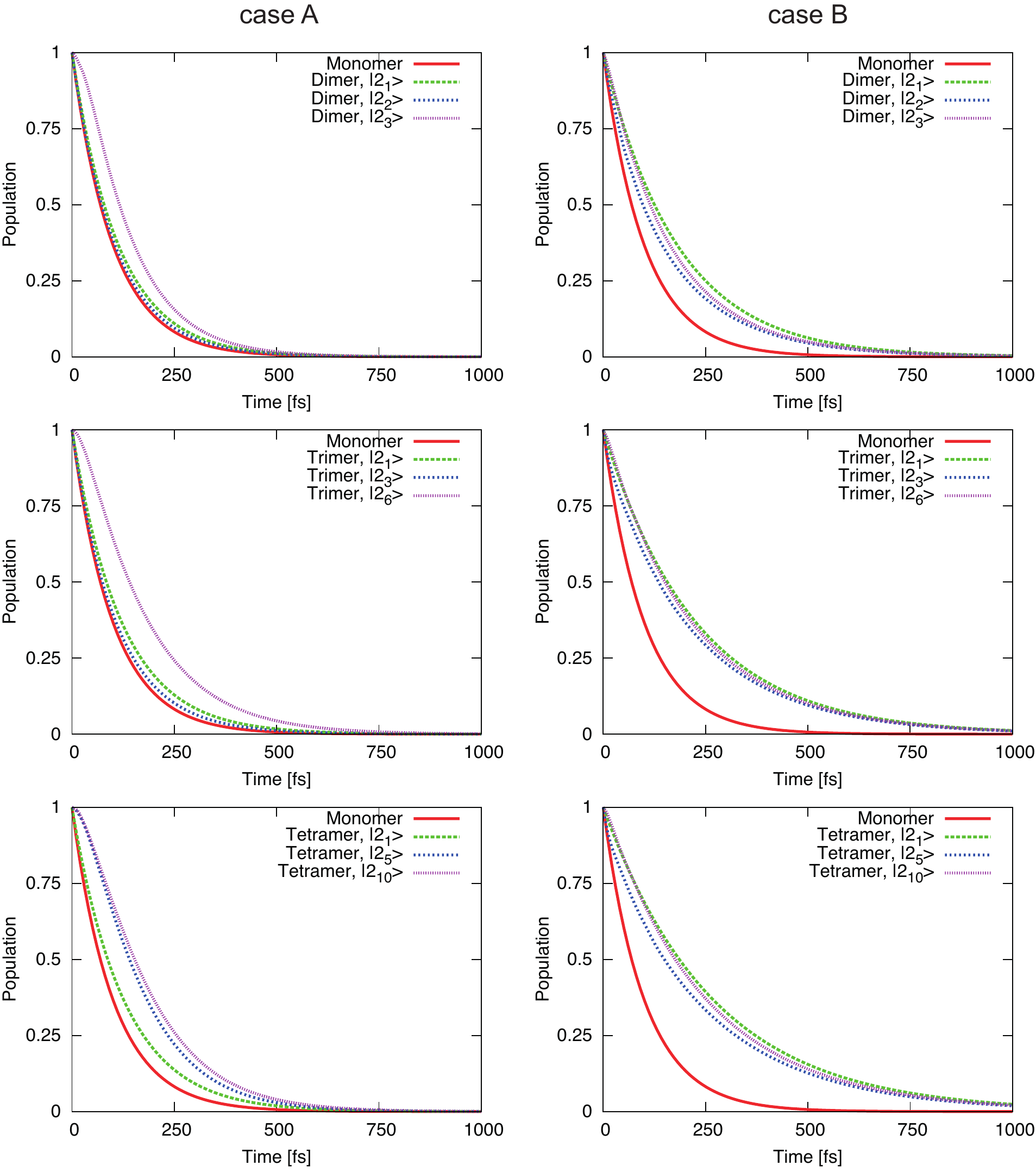}
\caption{Population dynamics in the field-free case after initial population of a specific two-exciton eigenstate. Shown is the total population of the two-exciton manifold in comparison with the monomer case.}
\label{fig:ic}
\end{center}
\end{figure}

Aggregates of three different sizes ($N=$2, 3, and 4 sites) will be studied for cases A and B (notation $NA/NB$). The one- and two-exciton eigenstates are shown in Fig. \ref{fig:level}. Clearly, for cases A representing a larger anharmonicity the locally double excited states are energetically separated from the delocalized double excitation. For instance, in the dimer we have for case A $\ket{2_1}=0.654\ket{11} +0.654\ket{22} +0.378\ket{12}$, $\ket{2_3}=-0.268\ket{11} -0.268\ket{22} +0.926\ket{12}$, i.e. the lowest/highest state has a dominant local/nonlocal character, whereas for case B we have $\ket{2_1}=0.537\ket{11} +0.537\ket{22} +0.650\ket{12}$ and 
$\ket{2_3}= -0.460\ket{11} -0.460\ket{22} +0.760\ket{12}$, i.e. local and nonlocal double excitation are mixed. Similar results are found for $N=3$ and 4.

\begin{figure}
\begin{center}
\includegraphics[width=\textwidth]{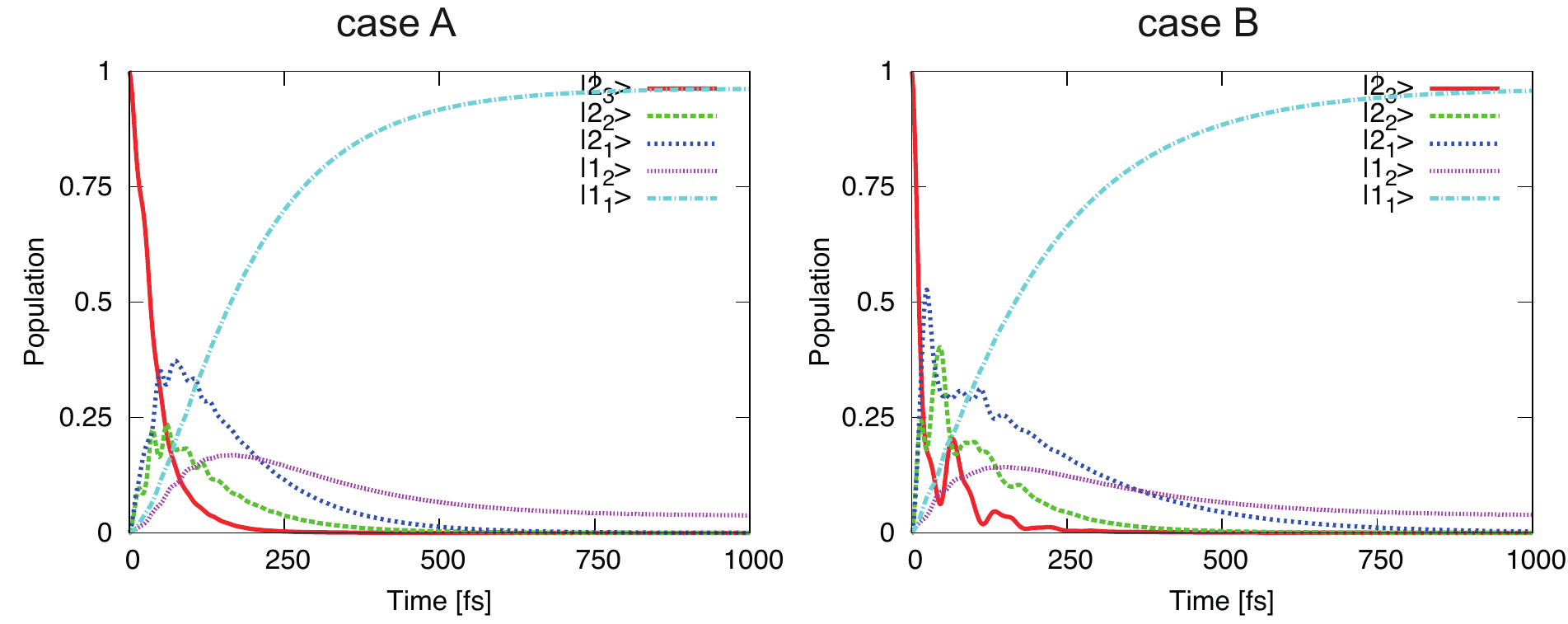}
\caption{Population dynamics of the dimer in the field-free case after initial population of the highest state of the two-exciton band. Shown are the populations of the one- and two-exciton eigenstates.}
\label{fig:intra}
\end{center}
\end{figure}
\begin{figure}
\begin{center}
\includegraphics[width=\textwidth]{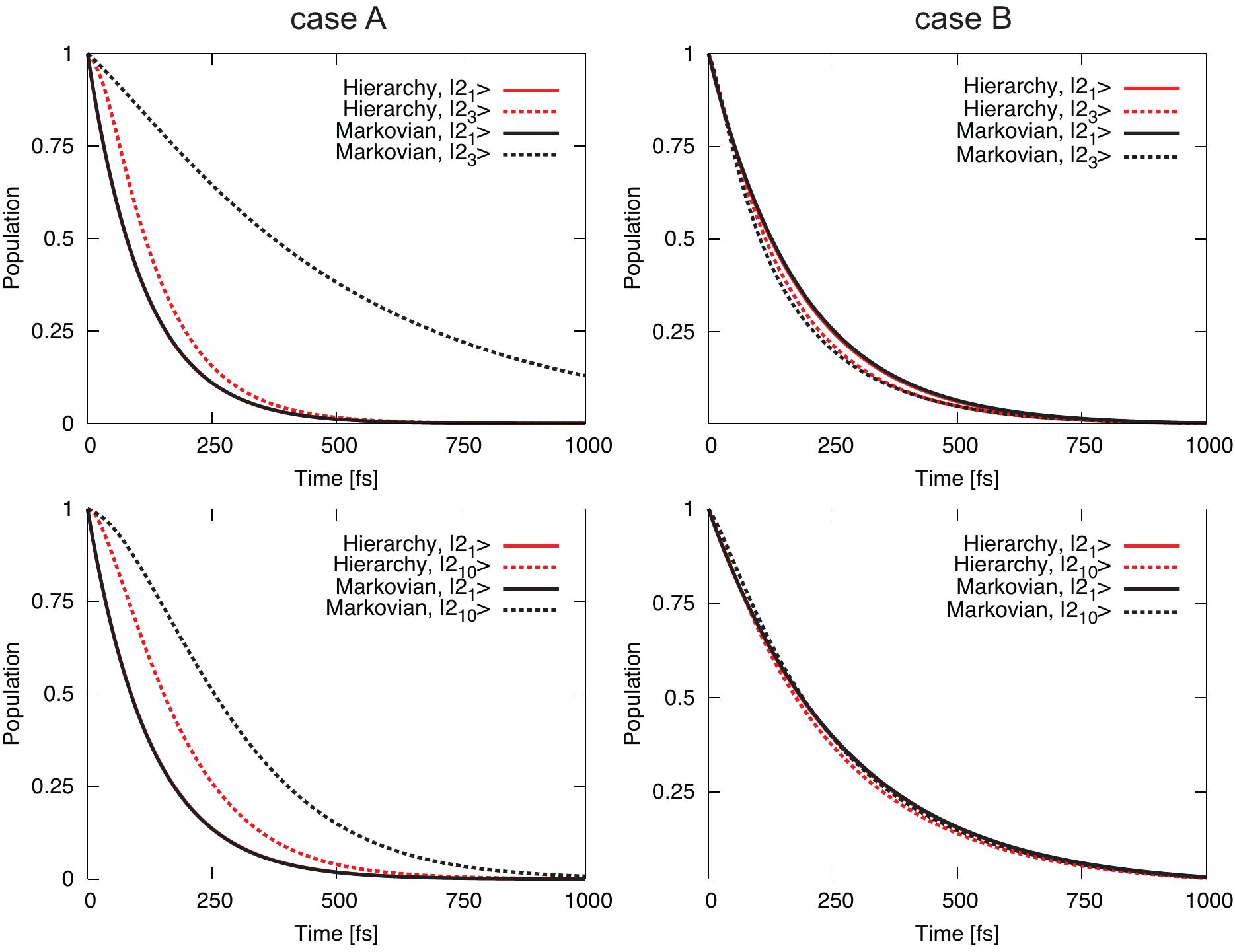}
\caption{Population dynamics in field-free case after initial population of the highest and lowest state of the two-exciton band (top row: dimer, bottom row: tetramer). Shown are the total populations of the  two-exciton bands for the HE and the Markovian case.}
\label{fig:free-comp1}
\end{center}
\end{figure}

\subsection{Population Dynamics in Field-Free Case} %------------------------------------
In Fig. \ref{fig:ic} we compare the internal conversion dynamics for the different models as a function of the initial preparation of the two-exciton manifold. Also shown is the decay of the monomer LDE state, which is always faster than that for the aggregate. Furthermore, it is found that the decay in case B is always slower as compared to case A, with the difference becoming more apparent upon increasing the aggregate size. This observation can be traced to the fact, that the mixing of LDE and NDE states in case B leads to a different intra-band dynamics and in particular to a slow-down of the inter-band dynamics as can be seen from Fig. \ref{fig:intra}. From this figure we notice that even though the decay of the initial state population is faster in case B, the population gets trapped for a longer time in state $\ket{2_1}$, i.e. at the bottom of the two-exciton band.
In case A the state $\ket{2_1}$ is of more local character as compared with case B and therefore inter-band relaxation is faster. Similar arguments apply to the aggregates with $N=3$ and 4.

\begin{figure}
\begin{center}
\includegraphics[width=\textwidth]{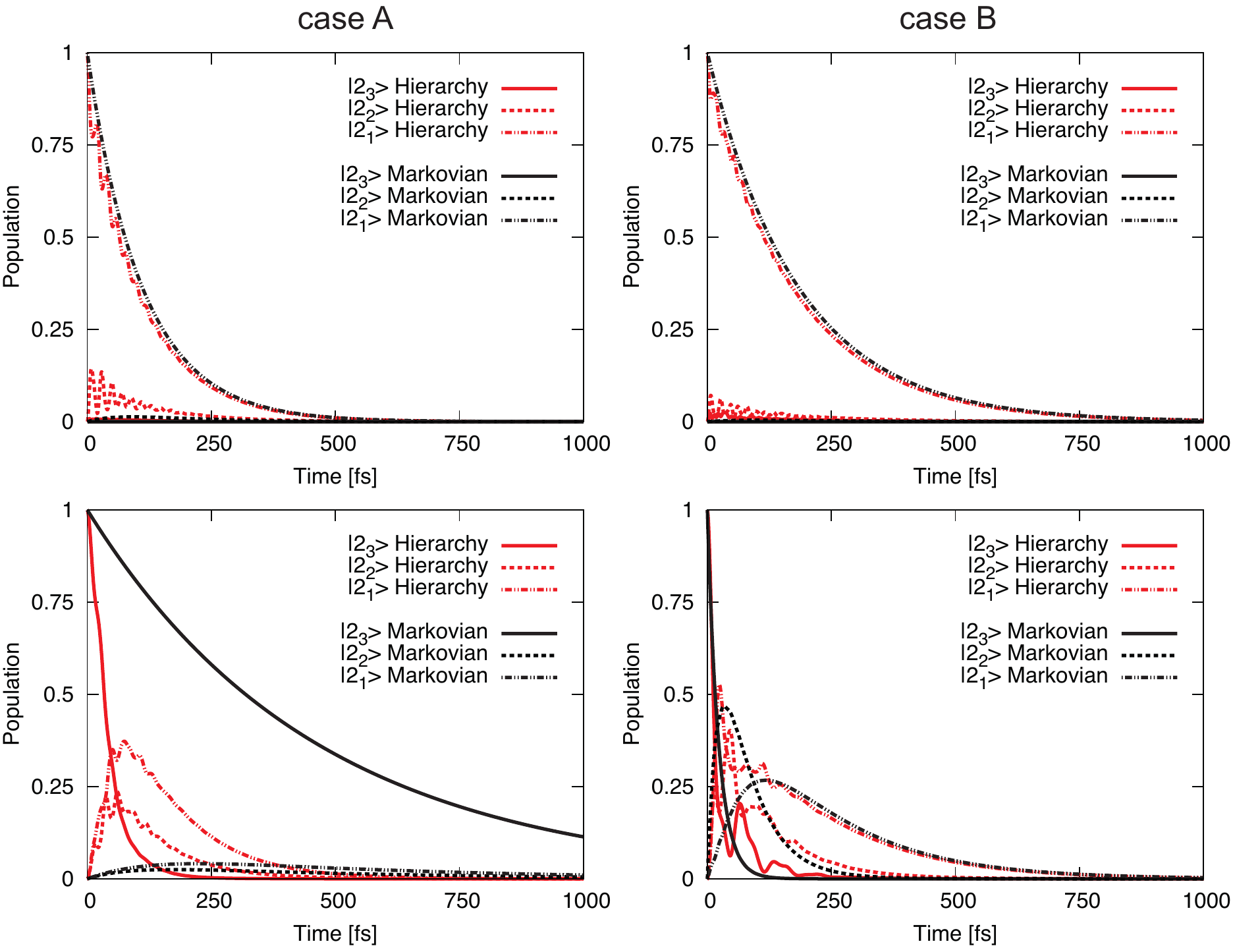}
\caption{Population dynamics of two-exciton eigenstates for 
dimer. Shown are the population in each state starting from the lowest (top row) and 
  highest (bottom row) state in the two-exciton band.}
\label{fig:free-comp2}
\end{center}
\end{figure}

So far we have presented results from the HE simulation. In Fig. \ref{fig:free-comp1} we show a comparison between HE and Markovian dynamics. In case B one hardly notices any difference in the dynamics, whereas case A shows considerable difference between the HE and the Markovian dynamics if the system is prepared in the uppermost two-exciton state. In order to scrutinize this effect we  have plotted the population dynamics of the two-exciton eigenstates in Fig. \ref{fig:free-comp2}. First, we notice that in general the populations from the HE are showing an oscillatory behavior, which is not present in the Markovian approximation. Despite this difference in details, the total two-exciton population is rather similar in the two limits. Only for case A and for initial preparation in state $\ket{2_3}$ we notice a marked deviation for the dimer considered in Fig. \ref{fig:free-comp2}. In the HE case state $\ket{2_3}$ relaxes rapidly and  states $\ket{2_2}$ and $\ket{2_1}$ become populated before the overall decay due to IC sets in. Notice that the energy gap between states $\ket{2_3}$ and $\ket{2_2}$ amounts to 2518 \cm. The spectral density for the system-bath interaction is composed of a low-frequency part with cut-off at 100 \cm and a high-frequency contribution at 1415 \cm. Hence relaxation within second order perturbation theory as implied in the Markovian, i.e. truncated HE, approach will be very inefficient due to the smallness of the spectral density. The HE, on the other hand, accounts for higher order effects and yields a rapid energy relaxation. Notice that in case of model B this energy gap is only 1203 \cm, i.e. within the frequency range covered by the spectral density. According to the argument given one would not expect a marked difference between HE and Markovian dynamics in case B, what is in line with the results shown in Fig. \ref{fig:free-comp1}. Inspecting Fig. \ref{fig:level} we note that a similar energy gap exists also for larger aggregates what leads to a corresponding behavior of the population dynamics in Fig. \ref{fig:free-comp1}. However, upon further increase of the aggregate size this gap closes and it can be expected that HE and Markovian dynamics behave rather similar at least from the total population point of view. In passing we note that ultrafast inter-band relaxation suppresses the effect of excitonic-polaron formation, which, however, will play a role in the one-exciton manifold (see, Ref. \cite{gelszinis11_245430}).

\subsection{Laser Control of Two-Exciton Dynamics} %---------------------------------
In the following we will investigate the possibility to control two-exciton dynamics with ultrashort shaped laser pulses. In particular we will ask the question whether one can transiently suppress EEA. Since EEA is a local process one might argue that a two-exciton wave packet, which is prepared in a way such that the contribution coming from LDE is minimized will lead to a slow down of EEA since the latter is mediated by intra-band relaxation processes mixing LDE and NDE states. For this purpose we will use the target state $\ket{1N}$ within the OCT scheme, i.e. the  state corresponding to the situation where the two NDEs have the largest separation in real space. The fastness of  the IC process puts some restriction on the pulse duration which has been set to $T=50$ fs. As mentioned in Sec. \ref{sec:oct} the field-optimization will be performed in the Markovian limit, starting with a very broad band pulse. The field is then used to generate the dynamics using the HE approach.

\begin{figure}
\begin{center}
\includegraphics[width=\textwidth]{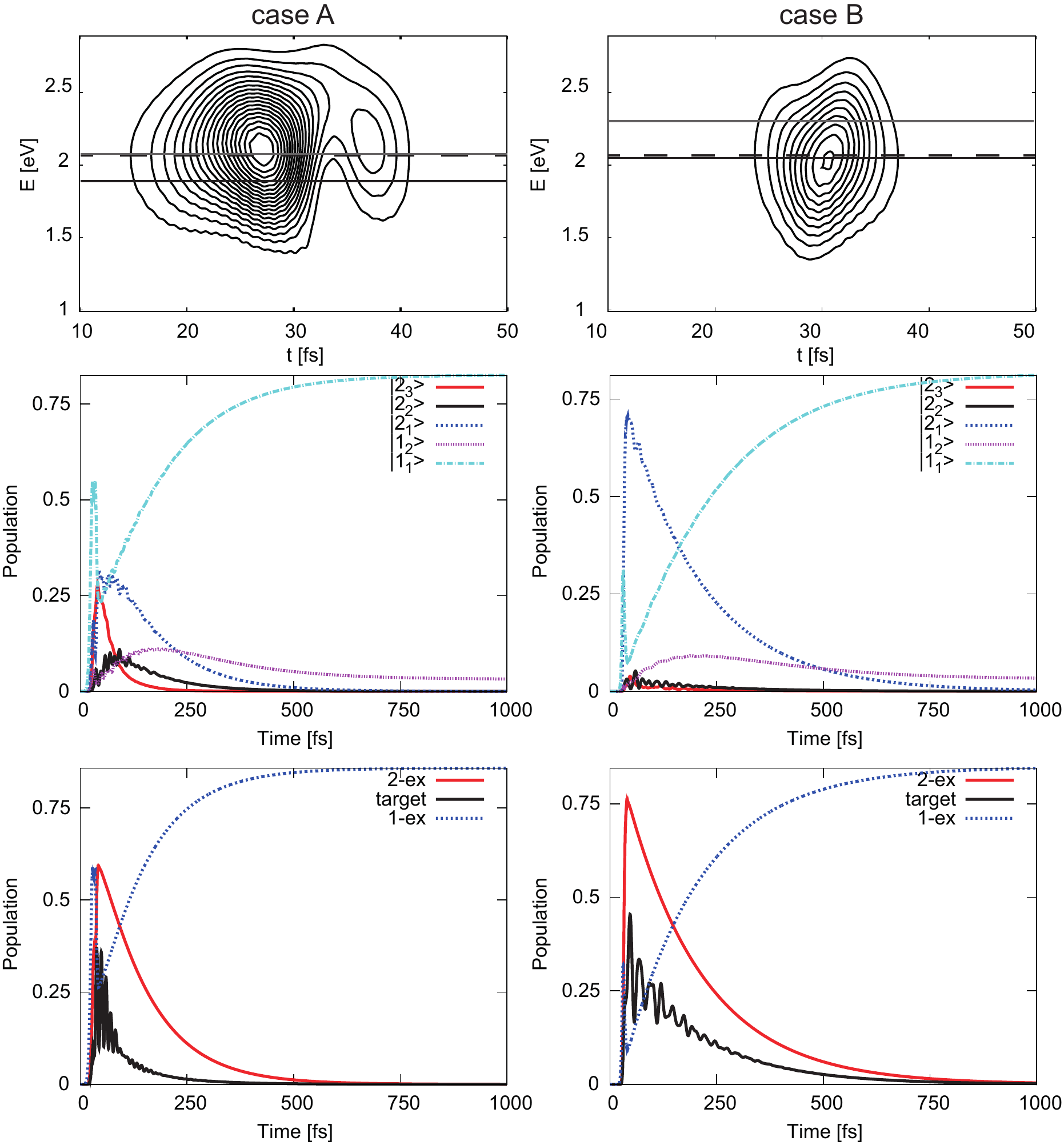}
\caption{Laser-driven dynamics of the two dimer models. The target state has been $\ket{12}$ and $I_0$ in \Eq{eq:octfunc} was set to 0.005 a.u.. Upper row: XFROG traces (Eq. (\ref{eq:xfrog})) of optimized field (20 (A) and 135 (B) iterations,  parameters for XFROG are $\tau=5$ fs, $\Delta=1$ fs). Contours are the equi-interval plots with 5\% increment of the peak height. The maximum field amplitude is 2.6 GV/m for case A and 3.1 GV/m for case B. The dashed line denotes the one-exciton resonance, the solid lines the one- to two-exciton resonances (colors as in Fig. \ref{fig:level}). Middle row: Population dynamics of one- and two-exciton eigenstates. Lower row: Dynamics of target state and total one- and two-exciton population.}
\label{fig:oct1}
\end{center}
\end{figure}

In Fig. \ref{fig:oct1} we show the optimized fields (XFROG) for cases A and B together with the population dynamics of target states as well as the exciton eigenstates for the dimer model. Since dipole transitions are possible between adjacent exciton bands only, the dynamics necessarily involves a two-step process. First, the one-exciton band is excited (transition frequency $\hbar \omega_{1_1,0}=2.06$ eV) and this process is followed by  a one- to two-exciton transition. In case A this way a superposition of states $\ket{2_1}$ and $\ket{2_3}$ is prepared (transition frequencies $\hbar \omega_{2_1,1_1}=1.88$ eV, $\hbar \omega_{2_3,1_1}=2.06$ eV), which, however, rapidly decays due to intra- and inter-band relaxation processes. For case B, where the LDE and NDE states are strongly mixed, the compromise found by the OCT equations  has been to prepare just state $\ket{2_1}$ which has a 42 \% overlap with the target state. Although state $\ket{2_3}$ would have had a larger overlap, its transition dipole matrix element is about a factor of 13 smaller for that case (transition frequencies: 
$\hbar \omega_{2_1,1_1}=2.04$ eV, $\hbar \omega_{2_3,1_1}=2.30$ eV). Similar to the scenario of the field free dynamics in Fig. \ref{fig:intra}, the overall decay of the two-exciton population is slower in case B as compared to case A. Finally, we notice that the maxima of the XFROG traces do not match with the bare exciton transitions in all cases, since superposition states are prepared by the broadband excitation. Further, exciton-vibrational coupling and the dynamic Stark shift will modify the bare excitonic resonances.

\begin{figure}
\begin{center}
\includegraphics[width=\textwidth]{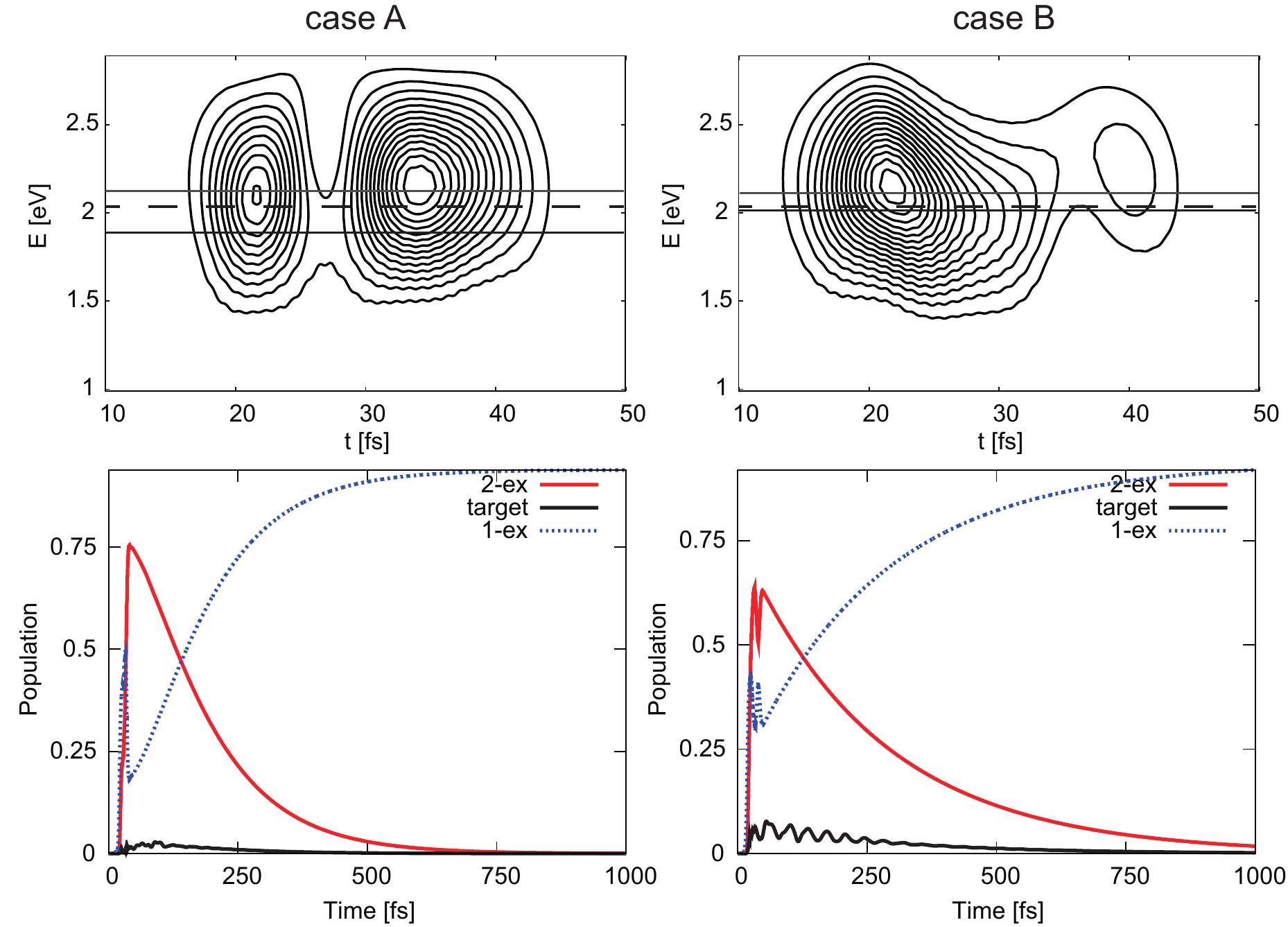}
\caption{Laser-driven dynamics of tetramer models. Upper row: XFROG traces (Eq. (\ref{eq:xfrog})) of optimized field (38 (A) and 30 (B) iterations, parameters for XFROG are $\tau=5$ fs, $\Delta=1$ fs). Contours are the equi-interval plots with 5\% increment of the peak height. The maximum field amplitude is 2.3 GV/m for case A and 2.4 GV/m for case B. The dashed line denotes the one-exciton resonance, the solid lines the one- to two-exciton resonances (colors as in Fig. \ref{fig:level}). Lower row: Dynamics of target state and total one- and two-exciton population.}
\label{fig:oct2}
\end{center}
\end{figure}

One might argue that in the dimer EEA will always be very effective since the NDE are localized at neighboring sites. This situation might change for larger aggregates. As an example we consider the tetramer model in Fig. \ref{fig:oct2}. The convergence of the OCT algorithm is rather slow in this case and for the given constraints only a small population of the target state can be reached. In both cases the pulse initially excites the lowest transition of the one-exciton band ($\hbar \omega_{1_1,0}=2.03$ eV). For case A the subsequent one- to two-exciton transition populates mainly state $\ket{2_5}$ ($\hbar \omega_{2_5,1_1}=2.12$ eV) which has the largest overlap with the target state (32 \%). In case B state $\ket{2_1}$ is dominantly populated ($\hbar \omega_{2_1,1_1}=2.01$ eV) although it has only a small overlap with the target state (3\%). However, as compared with state $\ket{2_4}$,  which has a 37 \% overlap with the target state, the transition dipole moment to state $\ket{2_1}$ is larger by a factor of about 5. Overall, the comparison between cases A and B resembles again Fig. \ref{fig:oct1}. 

Comparing the dimer and tetramer cases one notices that the initial hypothesis that the longer the aggregate the longer the time scale during which one can maintain a two-exciton population does not hold in general. Instead the difference between cases A and B points to the importance of mixing between LDE and NDE states within the two-exciton band. However, focussing on case B only there is the anticipated increase of the time-scale of transient two-exciton state population. In this case the optimized pulse populates the lowest state of the two-exciton manifold whose overlap with the target state decreases with increasing system sizes due to the ''dilution'' of the zero-order excitation states. For the same reason, however, the overlap of state $\ket{2_1}$ with LDE states decreases yielding a longer time scale for the two-exciton decay.

%The dynamics is dominated by the ultrafast relaxation processes, what raises the question whether  it really matters if an OCT or a transform limited pulse shape is used. To scrutinize this issue we have repeated the HE simulation with Gaussian pulses whose properties are derived from the respective optimized ones.
%
%The light field may contain 
%more than one pulse depending on the dynamics under investigation
%\begin{eqnarray}
%\label{eq:pulse}
%\svec{E}(t) = \sum_j \svec{e}_jE_j(t) 
%    \exp(i\svec{k}_j\cdot\svec{r} - i\omega_j t) + c.c.
%\end{eqnarray}
%where $\svec{e}_j$ is the polarization, 
%      $E_j(t)$ the envelop in time domain, 
%      $\svec{k}_j$ the wave vector, and
%      $\omega_j$ central frequency of the $j$th pulse.
%Here we assume the pulse is Gaussian having full width at half maximum (FWHM)
%$\delta_j$, i.e. $E_j(t) = T_j \exp(-4\ln 2(t-\tau_j)^2/\delta_j^2)$ with
%$T_j$ being the field strength and $\tau_j$ the time delay.
%
%%
%\begin{figure}
%\begin{center}
%\includegraphics[width=\textwidth]{fig8.eps}
%\caption{Population dynamics in field-free case after initial population of the highest and lowest state of the two-exciton band (top row: dimer, bottom row: tetramer). Shown are the total populations of the  two-exciton bands for the HE and the Markovian case.}
%\label{fig:comp1}
%\end{center}
%\end{figure}
%%
%
\begin{figure}
\begin{center}
\includegraphics[width=\textwidth]{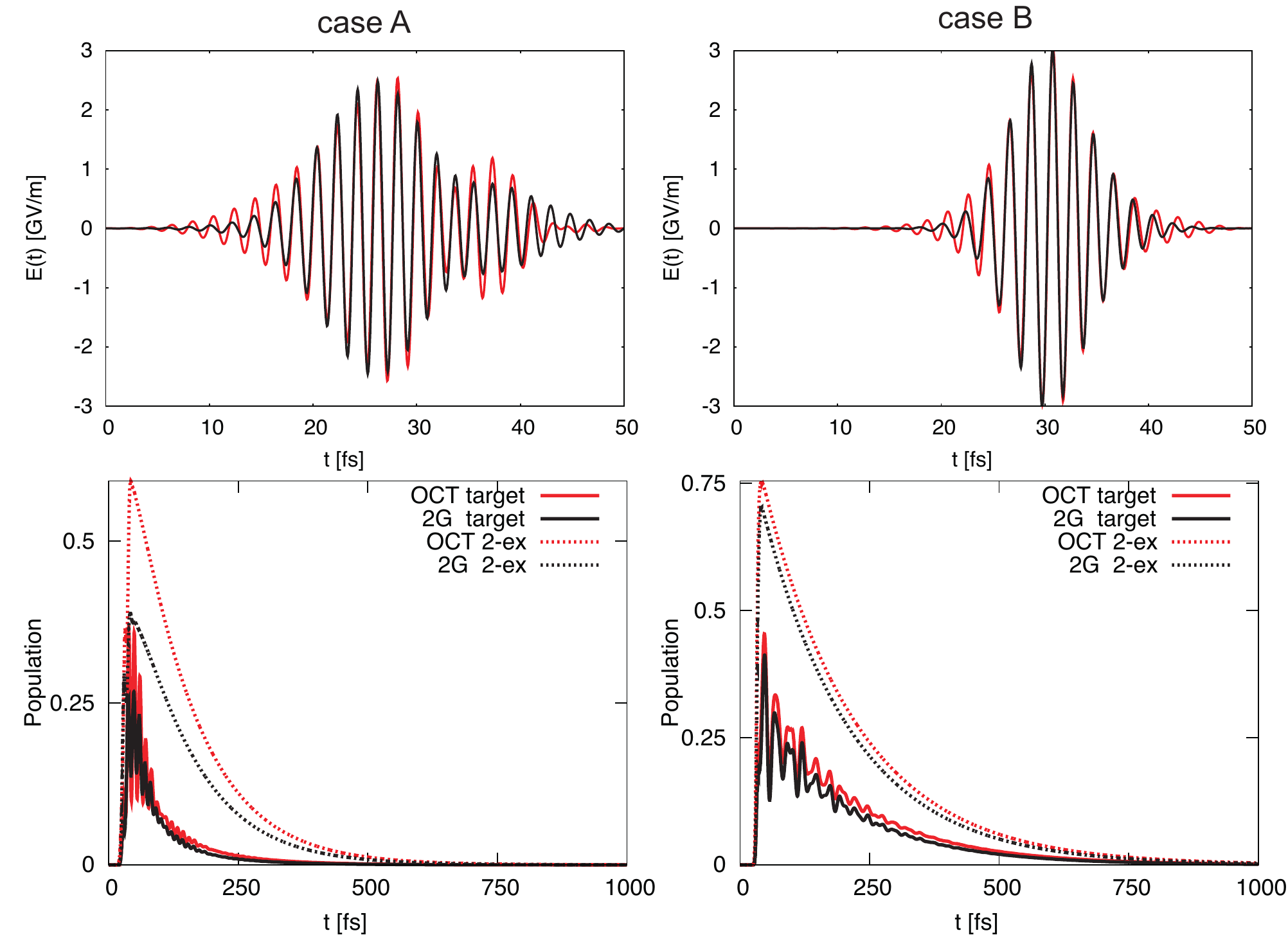}
\caption{Comparison of laser-driven dynamics for the dimer models using the pulses obtained from the OCT equations (red) and a fit of these pulses to two Gaussians (2G, black). The lower panels show the resulting target and two-exciton manifold populations.}
\label{fig:gaussian}
\end{center}
\end{figure}
Next we comment on the use of pulses derived by using the OCT equations and  Markovian dynamics within the HE scheme. Inspection of the different cases shows that the resulting population of the target state is comparable in the two cases, i.e. this procedure does not lead to a degradation of control for the given pulse shape. Needless to say that in line with the discussion of Fig. \ref{fig:free-comp1} the relaxation dynamics after the pulse is over differs. Of course, one might argue that the combination of OCT equations and HE might result in different pulses. However, in view of the difficulties arising from  the short life time of the two-exciton population this will have little relevance for the present discussion.

We need to emphasize that our model is limited to the two-exciton space. In principle for the given field intensities it is likely that higher exciton manifolds could be excited as well (for a study of intensity-dependent multi-exciton dynamics within a Bloch model, see Ref. \cite{richter07:075105}). The expected rapid decay of these states would lead to an additional channel for populating the two-exciton manifold, what could in principle influence the details of the dynamics. Needless to say, that this effect could be suppressed by reducing the field intensity, but  at the expense of the population of the two-exciton manifold.

Finally, the question arises whether all the details of the control pulses play a role or in other words how robust is the achieved population control and can the pulse be simplified. Exemplarily, in Fig. \ref{fig:gaussian} we show a comparison of the dynamics obtained for the OCT derived pulses (cf. Fig. \ref{fig:oct1}) and their fits to two Gaussian shaped pulses for  the dimer model. For case A we observe a strong sensitivity, i.e. the two-exciton population drops down considerably when using two Gaussian pulses. Apparently, the preparation of the $\ket{2_1}$ and $\ket{2_3}$ superposition state depends on the details of the control field. In contrast, in case B where dominantly the state 
$\ket{2_1}$ is prepared a simple Gaussian-shaped field is sufficient to achieve a control comparable to the optimized pulse.

\section{Summary} %---------------------------------
\label{sec:sum}
Two-exciton dynamics has been studied on the basis of a non-perturbative and non-Markovian hierarchy equation approach. The dissipative dynamics has been combined with optimal control theory for obtaining laser fields designed such as to trigger long-lived two-exciton state populations. Specific results have been obtained for short aggregates made of J-aggregate forming perylene bisimide dyes for which Frenkel exciton parameters are available \cite{ambrosek11_17649}. Establishing a two-exciton population one has to compete with the very efficient internal conversion, which is a consequence of the breakdown of the Born-Oppenheimer approximation. For the monomer this process takes place on a time scale of about 100 fs, which therefore sets the upper bound for laser control. The fact that two-exciton populations can be maintained on a longer time scale is due to the mixing between local  and non-local double excitations of the aggregate; the latter do not decay on an ultrafast time scale. The considered dyes support two possible double excitation states in the relevant energy range. Therefore, we considered two scenarios corresponding to small and substantial mixing between the two type of aggregate excitations. It turned out that the case of strong mixing allows for maintaining a  two-exciton population on a longer time scale (in the present cases about 1 ps as compared with 0.5 ps for the weak mixing case). This effect should be observable in a pump-probe experiment. At this point it should be noted that two-exciton populations in aggregates made of organic dyes like PBI have been observed for the first time only very recently \cite{wolter12}, what demonstrates feasibility of preparation and spectral identification.

The present investigation highlights the importance of zero-order state mixing within the two-exciton band for the inter-band transitions. The more diluted the zero-order states the more the decay is slowed down. However, this situation might change if static disorder is taken into account which will lead to a localization of the two excitations on different parts of the aggregate. 
\section*{Appendix: Derivation of the Hierarchy Equations of Motion} %-----------------------------------------------------
  \setcounter{equation}{0}%
  \renewcommand{\theequation}{A.\arabic{equation}}%

For the Caldeira-Leggett model of dissipation, Eq. (\ref{eq:hsb}), with  
$\hat{H}_{\rm S-B} = f(\hat{s}) g(\hat{b})$,
one can employ a stochastic procedure to derive the equation 
of the motion for the reduced density matrix of the relevant system $\hat{\rho}$
if the whole system is prepared as a factorized state, i.e.
$\hat{\rho}_{\rm tot}(0) = \hat{\rho}(0)\exp(-\beta \hat{H}_{\rm B})
           /\textrm{Tr}[\exp(-\beta \hat{H}_{\rm B})]$ \cite{shao04_5053,yan04_216}.
Within this approach, the influence of the bath is completely 
characterized by its response function and its effect is
described by a random force $\bar{g}(t)$ in the It\^o  stochastic differential equation
\begin{eqnarray}
\label{eq:drho}
\fl
i\hbar d \hat{\rho} =
[H_{s}+\bar{g}(t)f(\hat{s}),\rho_{s}] dt + \sqrt{\hbar} /2\left[
f(\hat{s}), \hat{\rho}\right]dw_{1,t} +i\sqrt{\hbar}
/2\left\{f(\hat{s}), \hat{\rho} \right\}dw_{2,t}^{*},
\end{eqnarray}
where $w_{1,t},\, w_{2,t}$ are two independent complex-valued Wiener processes 
and $w_{1,t}^\ast,\, w_{2,t}^\ast$ are their complex conjugates. In Eq. (\ref{eq:drho})
$\bar{g}(t)$, which plays the same role as the influence functional 
in the path integral treatment, can be regarded as a
stochastic field fully characterizing the influence of the environment. It is defined as
\begin{equation}
\label{eq:infl}
\bar{g}(t)=\sqrt{\hbar} \int_{0}^{t}\Big\{
\alpha_{R}(t-t') dw_{1,t'}^{*}+\alpha_{I}(t-t')dw_{2,t'} \Big \}.
\end{equation}
For simplicity, here it is assumed that the response function of the
bath is
\begin{eqnarray}
\label{eq:alp}
\alpha(t) = \sum_{k=1}^2 b_k e^{-\Omega_k t},
\end{eqnarray}
where $\Omega_1 = \Omega_2^*$ (cf. Section \ref{sec:timescale}). Note that the extension to multiple exponentials is straightforward. For a more general form of the response function consult Ref. \cite{yan04_216,ishizaki05:3131}. 

For the case of Eq. (\ref{eq:alp}) the stochastic force can be written as
\begin{eqnarray}
\bar{g}(t) = \sqrt{\hbar} \int^t_0 \exp[-\Omega_1 (t-\tau)] 
                 (\alpha_{1;+} dw_{1,\tau}^\ast + \alpha_{1;-} dw_{2,\tau}) \nonumber \\
            +\sqrt{\hbar} \int^t_0 \exp[-\Omega_2 (t-\tau)] 
                 (\alpha_{2;+} dw_{1}^\ast + \alpha_{2;-} dw_2) \nonumber \\
           \equiv \bar{g}_1(t) + \bar{g}_2(t).
\end{eqnarray}
Here $\alpha_{k;\pm}$ are defined as  $\alpha_{1;+} = (b_1 + b_2^\ast)/2$,   $\alpha_{1;-} = -i (b_1 - b_2^\ast)/2$,
 $\alpha_{2;+} = (b_1^\ast + b_2)/2$,  and  $\alpha_{2;-} = -i (-b_1^\ast + b_2)/2$.

Next we introduce the auxiliary reduced density matrices 
\begin{eqnarray}
\hat{\rho}_{m,n}(t) = \frac{\Re(\Omega_1)^{m}\Re(\Omega_2)^{n}}{\sqrt{m! n! (c^{\rm (H)}_0)^{m+n}}}
  M\left\{\bar{g}_{1}^{m}(t) \bar{g}_{2}^{n}(t) \hat{\rho}(t)\right\},
\end{eqnarray}
where $M\{\cdots\}$ denotes the ensemble average with respect to the noise and $c^{\rm (H)}_0 = \sqrt{|b_1 b_2|}$.
Notice that using a scaling like in the above equation has been suggested by YiJing Yan and co-workers~\cite{zhu11_5678,shi09_084105}. The present prefactor is slightly different from that suggested in Refs. \cite{zhu11_5678,shi09_084105}, where the numerator was set equal to one. This choice  will keep the terms in the same order size consistent in cases where the decay constants in the response functions are rather different.

Carrying out the stochastic average for the random variables 
in Eq.~(\ref{eq:drho}), elementary stochastic
calculus yields the differential equation for $\hat{\rho}_{mn}(t)$
\begin{eqnarray}
\label{rho:mn} 
\fl
i\hbar d\hat{\rho}_{m,n}(t)/dt =
	- i\hbar(m \Omega_1 + n \Omega_2)\hat{\rho}_{m,n}(t) 
	+ [H_{s},\hat{\rho}_{m,n}(t)] \nonumber \\
	+ \Re(\Omega_1)^{-1} \sqrt{(m+1)c^{\rm (H)}_0} [f(\hat{s}),\hat{\rho}_{m+1,n}(t)] \nonumber \\
  + \Re(\Omega_2)^{-1} \sqrt{(n+1)c^{\rm (H)}_0} [f(\hat{s}, \hat{\rho}_{m,n+1}(t)] \nonumber \\
	+ \hbar \Re(\Omega_1) \sqrt{m/c^{\rm (H)}_0}
      \Big(\alpha_{1;+} [f(\hat{s}), \hat{\rho}_{m-1,n}(t)] 
        - \alpha_{1;-}\{f(\hat{s}), \hat{\rho}_{m-1,n}(t)\} \Big) \nonumber \\
  + \hbar \Re(\Omega_2) \sqrt{n/c^{\rm (H)}_0} 
      \Big(\alpha_{2;+} [f(\hat{s}), \hat{\rho}_{m,n-1}(t)] 
        - \alpha_{2;-}\{f(\hat{s}), \hat{\rho}_{m,n-1}(t)\} \Big)
\end{eqnarray}
This equation needs to be solved for the 
initial conditions  $\hat{\rho}_{00}(0) = \hat{\rho}(0)$ and
$\hat{\rho}_{m, n}(0) = 0$ ($m \neq 0$ or $n \neq 0$). Application to the present system-bath model yields the equations of motion, Eq. (\ref{eq:EOM}).

\section*{Acknowledgemnt}
This work has been financially supported by the Deutsche Forschungsgemeinschaft through the Sfb 652.

\section*{References}
%\bibliography{/Users/ok/Documents/iDocuments/Bibliothek/BibTeX/all}
%\bibliography{all}

\end{document}